\documentclass[runningheads]{llncs}

\usepackage[T1]{fontenc}

\usepackage{cite}
\usepackage{amsmath,amssymb,amsfonts}
\usepackage{algorithmic}
\usepackage{graphicx}
\usepackage{textcomp}
\usepackage{multirow}
\usepackage{booktabs}
\usepackage{adjustbox}
\usepackage{url}
\usepackage{xcolor}
\usepackage[T1]{fontenc}
\usepackage{mathrsfs}
\usepackage[linesnumbered,ruled,vlined]{algorithm2e}
\usepackage{makecell}
\usepackage{float}
\usepackage{afterpage}
\usepackage{lipsum}
\usepackage{subfigure}
\usepackage{caption}
\usepackage{stfloats}
\usepackage{tcolorbox}
\usepackage{tabularx}
\usepackage{hyperref}
\hypersetup{hidelinks}
\usepackage{bbding}
\usepackage[marginal]{footmisc}
\newcommand{\repeatthanks}{\textsuperscript{\thefootnote}}

\begin{document}
\title{Dual prototype attentive graph network for cross-market recommendation}

\author{Li Fan\thanks{Both authors contribute equally to this research.}%
  \and Menglin Kong\repeatthanks%
  \and Yang Xiang%
  \and Chong Zhang%
  \and Chengtao Ji\Envelope%
}
\authorrunning{Li et al.}

\institute{Xi’an Jiaotong–Liverpool University, Suzhou, China\\
  \email{chengtao.ji@xjtlu.edu.cn}
}


\maketitle

\begin{abstract}
    Cross-market recommender systems (CMRS) aim to utilize historical data from mature markets to promote multinational products in emerging markets. However, existing CMRS approaches often overlook the potential for shared preferences among users in different markets, focusing primarily on modeling specific preferences within each market. In this paper, we argue that incorporating both market-specific and market-shared insights can enhance the generalizability and robustness of CMRS. We propose a novel approach called Dual Prototype Attentive Graph Network for Cross-Market Recommendation (DGRE) to address this. DGRE leverages prototypes based on graph representation learning from both items and users to capture market-specific and market-shared insights. Specifically, DGRE incorporates market-shared prototypes by clustering users from various markets to identify behavioural similarities and create market-shared user profiles. Additionally, it constructs item-side prototypes by aggregating item features within each market, providing valuable market-specific insights. We conduct extensive experiments to validate the effectiveness of DGRE on a real-world cross-market dataset, and the results show that considering both market-specific and market-sharing aspects in modelling can improve the generalization and robustness of CMRS.
\end{abstract}

\section{Introduction}
The development of e-commerce has prompted multinational companies to utilize data from multiple country markets, aiming to expand their market share and create sales opportunities \cite{li2019ddtcdr}. As a result, cross-market recommendation systems (CMRS) have emerged as a topic of interest. These systems analyze multiple markets to enhance recommendation strategies in emerging markets, leveraging user-item interaction data from well-established markets \cite{CoNet,MINDTL}. By conducting comprehensive analyses of user behavior and trends across different countries and regions, CMRS can provide users from diverse backgrounds with personalized and market-aware recommendations \cite{Contextual-Invariants}.

\begin{figure*}[t]
    \centering
    \includegraphics[width=\linewidth]{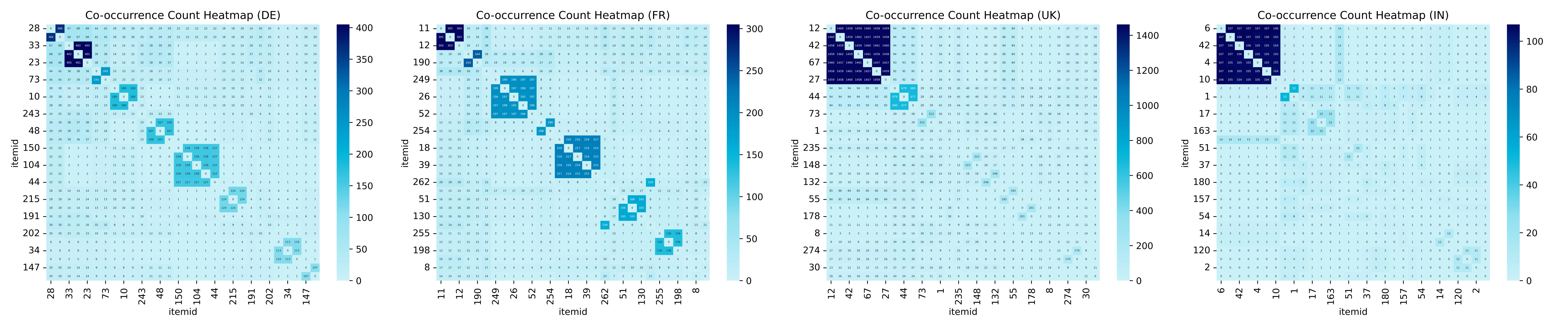}
    \caption{Heat map of item co-occurrence in different markets. The colour shade indicates the number of users interacting with the two items. In descending order, we sort the items in each marketplace by the number of co-occurrences and select the top 30 for displaying.}
    \vspace{-1\baselineskip}
    \label{fig:intro}
\end{figure*}

The challenges faced in CMRS include limited data from the target market and the influence of cultural differences on user preferences \cite{bonab2021crossmarket}. To address these challenges, researchers have introduced the concept of market adaptation and proposed various approaches \cite{CoNet,MINDTL}. Bonab et al. \cite{CoNet} introduced FOREC, a market adaptive algorithm based on meta-learning. FOREC is pre-trained on a dataset encompassing all markets, enabling quick adaptation to the target market. Bhargav et al. \cite{MINDTL} introduced the market-aware model, which represents each market using market embeddings to customize item representations for specific markets. However, existing CMRS primarily focus on modeling specific preferences in each market, overlooking the potential for shared preferences between users in different markets \cite{Item-Similarity-Mining}. For example, Japanese and French users with an interest in cooking may purchase cookbooks, kitchen tools, and ingredients. As shown in Figure \ref{fig:intro}, the item pairs with the highest number of common user interactions are all different across markets, reflecting the variability of user preferences across markets\cite{Leveraging-Behavioral-Heterogeneity}; on the other hand, the aggregation patterns of item pairs across markets are similar (e.g., DE and FR, UK and IN), which suggests that users from different markets often exhibit similar behaviors regardless of their nationality. We argue that incorporating market-specific and market-shared insights enhances the generalizability and robustness of CMRS.

In this paper, we introduce DGRE (Dual Prototype Attentive Graph Network for Cross-Market Recommendation), a novel approach that utilizes prototypes based on graph representation learning from both items and users to capture market-specific and market-shared insights. On the user side, we employ landmark-based graph clustering to group users from different markets into clusters, using these cluster representations as market-shared prototypes. On the item side, we create market-specific prototypes by aggregating item representations within each market, providing valuable insights into market-specific characteristics. The user representations obtained from a global user graph network encompassing users worldwide are refined by incorporating the market-shared prototypes. Similarly, item representations are derived from market-specific graphs using Graph Neural Network (GNN) techniques \cite{GraphSage, Wang-et-al-2019a} and enhanced by market-specific prototypes. This prototype-based attentive weighting ensures effective adaptation and personalization of recommendations for the target market\cite{Wu-et-al-2019b, Wang-et-al-2019h}. We evaluate our proposed method on the XMarket dataset \cite{bonab2021crossmarket}, comparing it to several competitive baselines. The experimental results demonstrate the superiority of considering both market-specific and market-shared aspects in CMRS. Our contributions can be summarized as follows:
\begin{itemize}
    \item The research paper introduces a novel approach, the Dual Prototype Attentive Graph Network (DGRE), for CMRS. DGRE leverages graph representation learning to capture market-specific and market-shared insights, improving the generalization and robustness of existing CMRS.
    \item DGRE incorporates market-shared prototypes by clustering users from various markets to identify behavioral similarities and create market-shared user profiles. Additionally, it constructs item-side prototypes by aggregating item features within each market, yielding valuable market-specific insights. 
    \item DGRE is designed as a model-agnostic framework, making it readily applicable to existing CMRS methods. Through experimental evaluations conducted on the XMarket dataset, DGRE showcases its effectiveness by outperforming several baseline methods. 
\end{itemize}

\section{Related Work}
\subsection{Cross Domain and Cross Market Recommendation}
\textbf{Cross Domain Recommendation.}  
In recent times, there have been two main approaches in the field of cross-domain recommendation models. The first approach focuses on improving knowledge transfer models, as exemplified by CoNet\cite{CoNet}, which utilizes cross-connections between feed-forward neural networks for knowledge transfer. MINDTL\cite{MINDTL}, on the other hand, combines Collaborative Filtering (CF) information from the target domain with rating patterns extracted from a cluster-level rating matrix in the source domain. DDTCDR\cite{li2019ddtcdr} introduces a novel technique that employs latent orthogonal mapping to extract user preferences across multiple domains while preserving relationships between users in different latent spaces.

The second approach involves connecting user preferences across domains~\cite{CDR4, CST, CDR6, CDR7, Contextual-Invariants, 2024c2dr}, which is closely related to our research. Methods like CST\cite{CST} use user embeddings learned in the source domain to initialize user embeddings in the target domain, ensuring alignment. Some other methods explicitly model the preference bridging process\cite{ CDR4, CDR6, CDR7, Contextual-Invariants}. Our research is particularly influenced by the concept of extracting contextual invariants\cite{Contextual-Invariants}. Based on these invariants (consistent behaviors across domains), we hypothesize the existence of common user behaviors or properties across different markets. Therefore, our approach explicitly models diverse human behaviors around the world to comprehensively capture the patterns of people.

\textbf{Cross Market Recommendation.} In contrast to cross-domain recommendation, cross-market recommendation (CMR) has received less attention in research. Initially, CMR gained traction in the field of music recommendation \cite{Exploring-Music-Diversity, Leveraging-Behavioral-Heterogeneity}. Ferwerda et al.\cite{Exploring-Music-Diversity} explored CMR with a focus on diversity across countries, while Roitero et al.\cite{Leveraging-Behavioral-Heterogeneity} delved into CMR within the music domain. They investigated the balance between local/single-market learning and a global model and introduced various training strategies.

Furthermore, the XMarket Dataset, created by Bonab et al.\cite{bonab2021crossmarket}, has been instrumental in CMR research. Bonab et al.\cite{bonab2021crossmarket} developed an approach related to meta-learning, enabling knowledge transfer from a source market to a target market by adjusting specific model layers. Subsequently, Cao et al.\cite{Item-Similarity-Mining} trained a model to understand item similarities within and across markets, leading to multi-market recommendations. More recently, Bhargav et al.\cite{Market-Aware-CMR} introduced a more efficient CMR approach by modelling markets without relying on meta-learning but by establishing embeddings as market-aware representations to effectively transfer market knowledge. In our work, we construct market-aware representations using graph embedding techniques.

\subsection{Graph Learning based Recommender System}
In recent years, rapid development has occurred in the field known as GLRS. GLRS utilizes advanced graph learning techniques to model user preferences, intentions, and item characteristics for making recommendations. Lately, there are three main approaches: Random walk approaches\cite{Bagci-and-Karagoz, Baluja-et-al., Jiang-et-al, liu2024clustering}, Graph embedding approaches\cite{Wang-et-al-2019h, Mikolov-et-al, Cen-et-al, Han-et-al., Hu-et-al-2018}, and Graph neural network approaches\cite{Fan-et-al, Xu-et-al, Wang-et-al-2019f, Wu-et-al-2019b, Cui-et-al, Ying-et-al, Wang-et-al-2019a, Wang-et-al-2019b, GraphSage, shu2025knowledge}.

For the random walk approach, typically, in a basic random walk-based Recommender System (RS)\cite{Baluja-et-al.}, an initial random walker traverses a given graph following predetermined transition probabilities for each step. The random walk process captures user-item preferences and interactions. The probability of reaching nodes after steps is used for ranking recommendations\cite{Baluja-et-al., Mikolov-et-al}. Graph embedding compresses nodes into low-dimensional representations, encoding graph structure for analyzing complex connections between nodes like users and items, giving rise to Graph Embedding-based Recommender Systems (GERS)\cite{Baluja-et-al., yin2024counterfactual}. It has three main implementations: graph factorization\cite{Wang-et-al-2019h}, graph distributed representation\cite{Mikolov-et-al, Cen-et-al}, and graph neural embedding (including graph auto-encoder)\cite{Han-et-al.,Hu-et-al-2018,Cen-et-al, shi2024two}. 

Lastly, Graph Neural Networks (GNNs) are neural networks used to analyse graph data. Due to their ability to effectively learn information representations, certain Recommender Systems (RS) have employed GNNs to tackle the key challenges posed by Graph Learning-based Recommender Systems (GLRS). Graph neural network approaches divide into 3 classes: Graph Attention network-based RS (GATRS)\cite{Fan-et-al, Xu-et-al, Wang-et-al-2019f, feng2025cofa}, Gated Graph Neural Network-based RS (GGNNRS)\cite{Wu-et-al-2019b, Cui-et-al}, and Graph Convolutional Network (including GraphSage\cite{GraphSage}) based RS (GCNRS)\cite{Ying-et-al, Wang-et-al-2019a, Wang-et-al-2019b}. Our work concentrates most on the GCNRS, especially GraphSage\cite{GraphSage}, which involves sampling neighboring vertices for each vertex in a graph and then using an aggregation function to combine the information from these neighbors. We utilize the GraphSage\cite{GraphSage} method to construct prototype representations for users and market-specific representations for items with fine-grained detail.

\section{PROBLEM FORMULATION \& PRELIMINARY}
Let $\mathbf{M} = \{\mathbb{M}_0, \mathbb{M}_1, \ldots, \mathbb{M}_t\}$ represent the set of markets, with $t$ denoting the total number of markets. For each market $\mathbb{M}_l$, we define the user set as $\mathbf{U_l} = \{\mathbf{u}_l^1, \mathbf{u}_l^2, \ldots \mathbf{u}_l^n \}$ and their corresponding embeddings are denoted as $\mathbf{P}^l = \{\mathbf{p_0}^l, \mathbf{p_1}^l, \\
\mathbf{p_2}^l, \ldots, \mathbf{p_n}^l\}$; the item set is $\mathbf{I}_l = \{\mathbf{i}_l^1, \mathbf{i}_l^2, \ldots, \mathbf{i}_l^m\}$ with corresponding embeddings $\mathbf{Q}^l = \{\mathbf{q_0}^l, \mathbf{q_1}^l, \mathbf{q_2}^l, \ldots, \mathbf{q_m}^l\}$. Here, $n$ represents the number of users, and $m$ represents the number of items in that particular market. It's important to note that in the context of CMRS, we assume that items are consistent across different markets, but users do not overlap among different countries. The primary objective of CMRS, specifically for the target market $\mathbb{M}_l$, is to go beyond utilizing data solely from that market and incorporate data from a set of other parallel markets, denoted as $\mathbf{M}_l \subseteq \mathbf{M} - \{\mathbb{M}_l\}$. By leveraging data from these parallel markets, CMRS aims to enhance the quality of recommendations for users in the target market. In essence, DGRE utilizes user and item embeddings to derive market-specific prototypes denoted as $\mathbf{O} = \{\mathbf{o_0}, \mathbf{o_1}, \ldots, \mathbf{o_t}\}$ and user-behavior prototypes $\mathbf{B} = \{\mathbf{b_1}, \mathbf{b_2}, \ldots, \mathbf{b_k}\}$, improving generalization and robustness in existing CMRS by capturing both market-shared and market-specific aspects.

\section{Methodology}
\begin{figure*}[t]
    \centering
    \includegraphics[width=\linewidth]{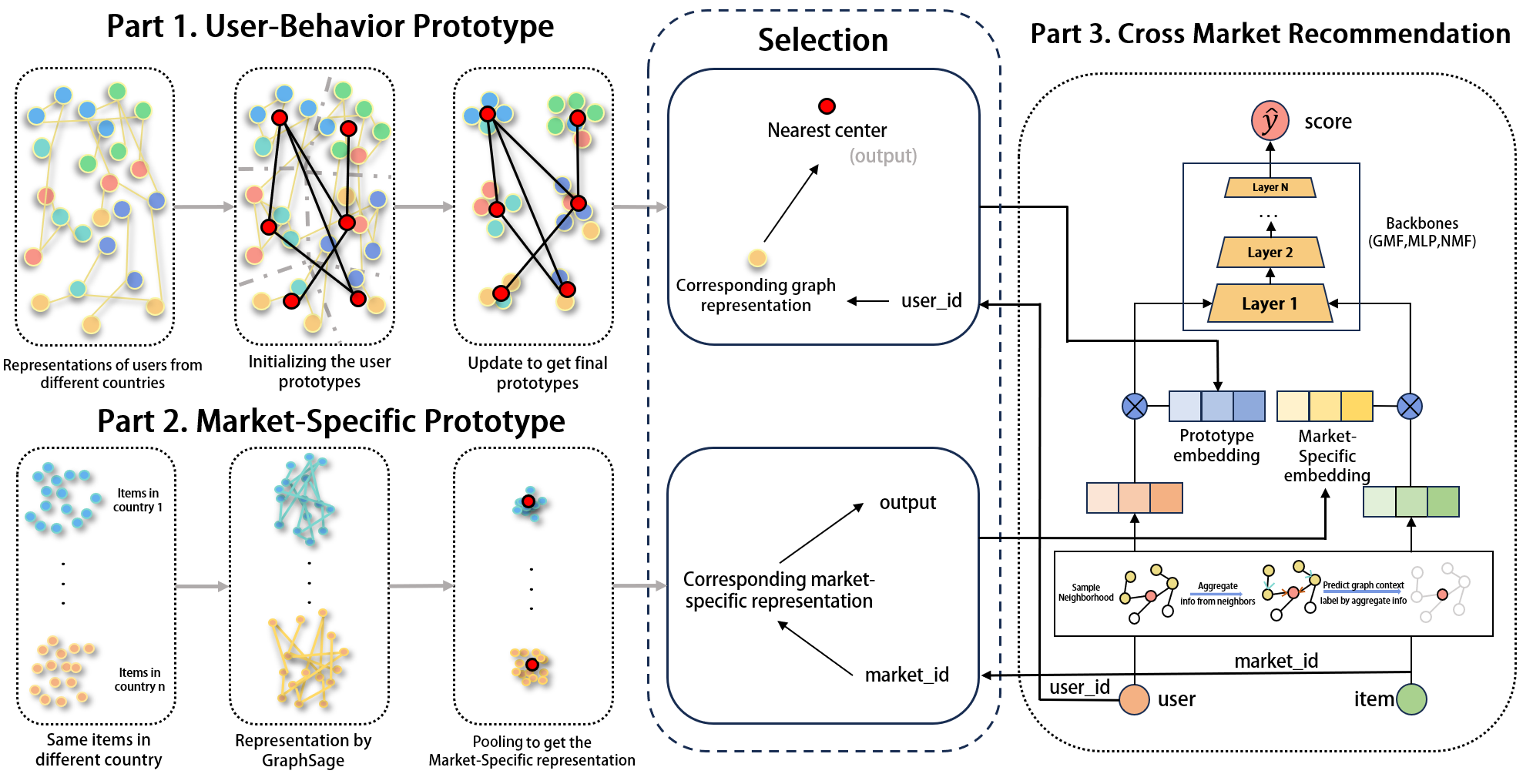}
    \caption{The framework of DGRE}
    \vspace{-1.5\baselineskip}
    \label{fig:your-label}
\end{figure*}

Our approach, shown in Figure \ref{fig:your-label}, focuses on obtaining two types of embeddings: market-shared prototypes and market-specific prototypes. For market-shared prototypes (section \ref{3.1}), we create a global user graph embedding from users across all markets and identify behavior patterns through landmark-based graph clustering. For market-specific prototypes (section \ref{3.2}), we construct item graphs for each market. We then pool each market's item graph to generate a prototype that reflects the market's characteristics. Our comprehensive Cross-Market Recommendation (CMR) framework combines these prototypes in the CMRS model (section \ref{3.3}).

\subsection{User-Behavior Prototypes}
\label{3.1}
In this section, DGRE creates user-behavior prototype representations that aim to classify users across all markets and enrich the expression of the interest of different kinds of users.

\subsubsection{Global User Embedding}
\label{3.1.2}
To start, we establish a global input graph that encompasses users from all markets, denoted as $G_{g} = (V_g, E_g)$. Here, $V_g$ represents the set of nodes (users across all markets), and $E_g$ represents the edges connecting two nodes (users) when they interact with at least two common items. 

Then, we employ GraphSage\cite{GraphSage}, a widely used method, on the global input graph. The core of our embedding algorithm is a convolution on the nodes' local neighborhood. To be more specific, it learns the embeddings of each node by updating the embedding \(e_{v}^{(k)}\), which a node \(v\) at layer \(k\) by aggregating embeddings from its neighbors \(N(v)\) and combining this with \(v\)'s current embedding, formalized as:

\begin{equation}
\begin{aligned}
e_{v}^{(k)} = \sigma\left(W^{(k)} \cdot \right. \quad \left. \text{AGGREGATE}\left(\{e_{v}^{(k-1)}\} \cup \{e_{u}^{(k-1)} : u \in N(v)\}\right)\right)
\end{aligned}
\end{equation}

where \(\sigma\) is a non-linear activation function, \(W^{(k)}\) is the layer-specific weight matrix, and \(\text{AGGREGATE}\) is a function that merges neighbor embeddings.

This iterative aggregation process across multiple layers enables embedding to capture node features and the context of the structural graph.

\subsubsection{User-Behavior Prototyping}
The goal of prototyping in the global user graph is to identify a set of informative structural representatives of various human behaviors. Inspired by the success of RDSA \cite{xiang2025rdsa} in the graph clustering task by selecting the most informative nodes as landmarks and using landmarks to optimize each substructure, we introduce a similar strategy to get the most informative nodes in each user-behavior class as the prototype for this class of users. Different from directly using RDSA, to fully capture the diversity of user interactions and sub-structures within the graph, we use modularity maximization\cite {newman2006finding} to pre-segregate users into different communities, which is the foundation for introducing the RDSA strategy. However, the traditional modularity maximization problem is an NP-hard problem \cite{newman2006finding}. Recent research has introduced soft assignment to replace the hard assignment in the modularity maximization problem by updating the soft assignment matrix \cite{DMoN}. Here, we calculate the cosine similarity matrix of embeddings $e$ as the soft assignment, formalized as:
\begin{equation}
    \begin{aligned}
        \boldsymbol{C}_{i,j} = \frac{e_i \cdot e_j}{\|e_i\| \|e_j\|},
    \end{aligned}
\end{equation}
where $\boldsymbol{C}_{i,j}$ is the cosine similarity between user $i$ and user $j$, and $e_i$ and $e_j$ are the embeddings of user $i$ and user $j$ respectively. Then, we define a set of prototypes as $\mathbf{B} = \{\mathbf{b_1}, \mathbf{b_2}, \ldots, \mathbf{b_k}\}$, where each prototype $\mathbf{b_k}$ aims to succinctly represent a sub-graph of similar user embeddings. In order to select the most representative user-behaviors as prototypes, we chose $k$ most respective nodes within each community based on its connectivity assess via modularity. The prototype selection process can be formulated as follows:
\begin{equation}
    \begin{aligned}
        \mathbf{B} = \underset{\mathbf{B} \subseteq \mathbf{U}, |U|=k}{\arg\max} \sum_{i} \sum_j\left(A_{i j}-\frac{d_i d_j}{2 m}\right) C(i, j),
    \end{aligned}
\end{equation}
where $A_{i j}$ is the adjacency matrix of the graph, $d_i$ and $d_j$ is the degree of user node $i$ and $j$, $m$ is the total number of edges in the graph, and $C(i, j)$ is the cosine similarity matrix of embeddings $e$. Individual user behavior is divided into its corresponding prototype by assessing the similarity of its representation to different prototypes. For each user-behavior, we construct a soft assignment matrix $W \in \mathbb{R}^{n_i \times k}$, where the $j^{th}$ entry of the $z^{th}$ column. Element $W(j, z)$ is the probability calculated based on the t-student distribution, which denotes the likelihood that the $j{th}$ user-behavior can be represented by the $k^{th}$ prototype $b_k$. The probability is influenced by the user embedding's proximity to prototypes. Specifically, the soft assignment $W(j, k)$ is given by:
\begin{equation}
W(j, k) = \frac{(1 + \|e_i(j, :) - b_k \|^2 / \alpha)^{-\frac{\alpha + 1}{2}}}{\sum_{k'}{(1 + \|e_i(j, :) - b_{k'} \|^2 / \alpha)^{-\frac{\alpha + 1}{2}}}}.
\end{equation}
Although we have selected prototypes for each user-behavior, we still need to consider the outliers and hard samples in the user-behavior class. In order to reduce the impact of outliers and hard samples, we introduce a self-shapen version of the t-student distribution $\tilde{W}$:
\begin{equation}
    \begin{aligned}
        \tilde{W}(j, k) = \frac{W^{2}(j,k)/\sum_{n}W(n,k)}{\sum_{k^{\prime}}[W^{2}(j, k^{\prime})/\sum_{n}W(n, k^{\prime})]}.
    \end{aligned}
\end{equation}
Then, we minimize the KL divergence between the $W$ and $\tilde{W}$ to shape the likelihood of user-behavior to the prototype during the training. Since the reshaping mechanism can reduce the impact of outliers and hard samples by capturing the diversity of user interactions and sub-structures within the graph, enabling efficient manipulation and interpretation of the graph's structural intricacies, this process can robustly get various prototypes in the global market.

\subsection{Market-Specific Prototypes}
\label{3.2}
In this phase, we aim to generate market-specific prototypes for each market.

\subsubsection{Embedding for Different Country Item Graph}
\label{3.2.2}
First of all, we generate item graphs for different countries independently. Then, for each item graph, we initiate the process by constructing an input graph, denoted as $G_{\mathbb{M}_{l}} = (V_l, E_{l})$ for market $\mathbb{M}_l$. Here, $V_{\mathbb{M}_{l}}$ represents the set of items within market $\mathbb{M}_{l}$, and $E_{\mathbb{M}_{l}}$ signifies the edges connecting at least two same users in market $\mathbb{M}_{l}$. Then, we generate the embeddings of nodes in the item graph similar to Section \ref{3.1.2}.

\subsubsection{Market-Specific Graph Pooling}
\label{3.2.3}
Following the generation of embeddings for nodes in different market-specific item graphs, we need to encode each item graph into a prototype. To achieve this, we introduce a graph pooling based on a selected node set for each market to get its market-specific prototype. The first step is to get the selected node set for different markets. The traditional methods to select vertices are based on the neural estimation of Mutual Information (MI)\cite{belghazi2018mutual}. However, existing algorithms are hard to accurately and efficiently approximate the mutual information. To solve this problem, we introduce the measurement from VIPool\cite{maosen2020vipool} which uses GAN-like divergence to measure the mutual information.
Firstly, to calculate the mutual information between the items and their neighborhood, we define the one-hop neighborhood of item $i$ as $N_{i}$. Then, we can get the aggregation of the neighborhood embeddings:
\begin{equation}
    \begin{aligned}
        q_{N_{i}} = \frac{1}{|N_{i}|} \sum_{j \in N_{i}} q_{j},
    \end{aligned}
\end{equation}
where $q_{j}$ is the embedding of item $j$ in the neighborhood of item $i$. Then, by VIPool\cite{maosen2020vipool}, we introduce the vertex selection process to select a set of items in the market that can represent the whole market. In a specific market $\mathbb{M}_{l}$, define the number of items to select as $k$, the set of most representative items as $S=\{s_1, s_2, \ldots, s_k\}$, the mutual information between the item $i$ and its neighborhood as $I_{GAN}(i)$. VIPool uses the GAN-like divergence to measure the mutual information of the vertex-neighborhood to achieve more flexibility and convenience in optimization:
\begin{equation}
    \begin{aligned}
        I_{GAN}(i, N_i)=\mathbb{E}_{P(i, N_i)}\left[\log \sigma\left(T\left(q_i, q_{N_i}\right)\right)\right]+\mathbb{E}_{P(i) \otimes P(N_i)}\left[\log \left(1-\sigma\left(T\left(q_i, q_{N_i}\right)\right)\right)\right],
    \end{aligned}
\end{equation}
where $\sigma$ is the sigmoid function, $T$ is a neural network that reflects the dependence between $q_i$ and $q_{N_i}$, $\mathbb{E}$ is the expectation. Through the mutual information between the items and their neighborhoods, we can select the set of items $S$ using greedy algorithm to maximize the mutual information between the prototypes and their neighborhood items through the following objective:
\begin{equation}
    \begin{aligned}
        \underset{S \in \mathbb{M}_{l}}{\max} \sum_{v \in S} I_{GAN}(v, N_v).
    \end{aligned}
\end{equation}
Once we have selected a set of items with the highest mutual information with their neighborhood, we can pool them into a single prototype that represents the whole market. To achieve this, we employ a weighted aggregation of the embeddings of the selected items to generate the market-specific prototype $c_{l}$ for market $\mathbb{M}_l$:
\begin{equation}
    \begin{aligned}
        c_{l} = \frac{\sum_{v \in S} I_{GAN}\left(v, N_v\right) \cdot q_v}{\sum_{v \in S} I_{GAN}\left(v, N_v\right)}.
    \end{aligned}
\end{equation}

\subsection{Model Training}
\label{3.3}
In the preceding section, we acquire market-specific prototypes for various markets (e.g., $\mathbf{o_{jp}, o_{ca}, ..., o_{us}}$) and market-shared prototypes for all users (e.g., $\mathbf{b_1}, \mathbf{b_2}, \ldots, \mathbf{b_k}$). Now, our objective is to integrate these two types of prototypes with the core recommendation architectures (namely, GMF, MLP, NMF) and basic embedding of the user (e.g., $\mathbf{p_i}^{ca}, \mathbf{p_i}^{us}, ...$) and item (e.g., $\mathbf{q_i}^{jp}, \mathbf{q_i}^{mx}, ...$) to train DGRE. In this section, we will use the symbol $\sigma$ to represent the activation functions, such as ReLU and Sigmoid.

\begin{itemize}
    \item \textbf{DGRE-GMF:} For a user $\mathbf{u}_i$ in market $l$ and item $\mathbf{i}_j$, we have user embedding $\mathbf{p_i}^l$, market-specific prototype $\mathbf{o_{l}}$, item embedding $\mathbf{q_{j}}^l$, and market-shared embedding $\mathbf{b_K}$. We calculate the prediction $\hat{y}_{\mathbf{u}_i,\mathbf{i}_j}$ as follows:
    \begin{equation} \hat{y}_{\mathbf{u}_i,\mathbf{i}_j} = \sigma(\mathbf{h}^{T}((\mathbf{\mathbf{p_i}^l}\odot \mathbf{b_K})\odot(\mathbf{o_{l}}\odot \mathbf{q_{j}}^l)) )). \end{equation}

    \item \textbf{DGRE-MLP:} The multi-layer perceptron (MLP) employs a fully-connected network with $L$ layers, where:
    \begin{equation}m_{0} = \left(\begin{matrix}
         \mathbf{\mathbf{p_i}^l}\odot \mathbf{b_K}\\
         \mathbf{o_{l}}\odot \mathbf{\mathbf{q_{j}}^l}
    \end{matrix}\right),\end{equation}
    \begin{equation}m_{L-1} = \sigma(W_{L}\sigma(...\sigma(W_{1}m_{0} + b_{1})) + b_{L}),\end{equation}
    \begin{equation}\hat{y}_{\mathbf{u}_i,\mathbf{i}_j} = \sigma(\mathbf{h}^{T}m_{L-1}).\end{equation}

    \item \textbf{DGRE-NMF:} Neural matrix factorization (NMF) combines elements of both MLP and GMF. Given $\mathbf{\mathbf{p_i}^l}^{MLP}$, $\mathbf{\mathbf{q_{j}}^l}^{MLP}$ as the pre-train parameters from the MLP in DGRE-MLP, and $\mathbf{\mathbf{p_i}^l}^{GMF}$, $\mathbf{\mathbf{q_{j}}^l}^{GMF}$ as the pre-train parameters from the GMF in DGRE-GMF, the NMF model computes the score as follows:
    \begin{equation}m_{0} = \left(\begin{matrix}
         \mathbf{\mathbf{p_i}^l}^{MLP}\odot \mathbf{b_K}\\
         \mathbf{o_{l}}\odot \mathbf{\mathbf{q_{j}}^l}^{MLP}
    \end{matrix}\right),\end{equation}
    \begin{equation}m_{MLP} = \sigma(W_{L} \sigma(...\sigma(W_{1}m_{0} + b_{1})) + b_{L}),\end{equation}
    \begin{equation}m_{GMF} = (\mathbf{\mathbf{p_i}^l}^{GMF}\odot \mathbf{b_K})\odot(\mathbf{o_{l}}\odot \mathbf{\mathbf{q_{j}}^l}^{GMF}),\end{equation}
    \begin{equation}\hat{y}_{\mathbf{u}_i,\mathbf{i}_j} = \sigma(\mathbf{h}^{T}\left(
         m_{GMF} ||
         m_{MLP}
    \right)).\end{equation}
\end{itemize}
These models are trained in a manner similar to the basic models (i.e., GMF, MLP, NMF), incorporating dual graph representation embeddings to enhance information on both the user and item sides.

\section{Experimental Evaluation}
\subsection{Experimental Setup}
In our experiments, we conduct two sets: "global experiments" and ablation studies to analyze Collaborative Market Recommendation (CMR) models\cite{Market-Aware-CMR}. In the "global experiment," we train these models collectively using data from all markets\cite{Market-Aware-CMR}. These experiments provide insights into model effectiveness and comparative performance.

\textbf{Dataset.} Our evaluations utilize the widely recognized XMarket dataset\cite{bonab2021crossmarket}, derived from a comprehensive Amazon dataset spanning multiple countries. Due to its size and relevance, we concentrate on the 'Electronics' subset. To ensure data quality, we apply Bonab et al.'s recommendation\cite {bonab2021crossmarket} and filter items and users with fewer than five interactions. Notice that the 'us' market, with the highest user-item interactions, is treated as a source market, not a target market.

\textbf{Compared Methods.} To assess the effectiveness of our models, we compare them against the following baseline models for each target market:
\begin{itemize}
\item \textbf{GMF, MLP, NMF:} Generalized Matrix Factorization (GMF), Multi-Layer Perceptron (MLP), and Neural Matrix Factorization (NMF) models\cite{NMF}, trained across all allowed source markets.
\item \textbf{Meta-learning Models:} We train these models using all market data as auxiliary markets.
    \begin{itemize}
    \item \textbf{MAML\cite{bonab2021crossmarket,MAML}:} This model is based on Model-Agnostic Meta-Learning (MAML) and is initialized with weights from a pre-trained NMF model.
    \item \textbf{NMF-FOREC\cite{bonab2021crossmarket}}: This model employs the pre-trained MAML model, initially freezing specific neural network layers and then fine-tunes on the target market.
    \end{itemize}
\item \textbf{Market-Aware Models\cite{Market-Aware-CMR}:} These models, proposed by Bhargav et al.\cite{Market-Aware-CMR}, utilize GMF, MLP, and NMF as backbones. They incorporate market-aware representations (e.g., one-hot encoding to distinguish markets) for training a CMR model. These models are denoted with the 'MA-' prefix in their names.
\end{itemize}

\textbf{Settings for Ablation Study.} we maintain a consistent backbone, GMF, in all ablation study experiments to ensure controlled variables during evaluation.

\textbf{Model Hyperparameters.} We configure the model parameters based on the specifications outlined in \cite{bonab2021crossmarket}. 
In the case of MAML models, we define a fast learning rate $\beta$ = 0.1 and use 20 shots for training, following the approach in \cite{Market-Aware-CMR}. Additionally, in the global experiment, we set the number of cluster centers $k$ to 10 in the landmark-based graph clustering process for the embedding of market-shared prototypes.

\textbf{Evaluation Metrics.} We evaluate our models using widely accepted metrics in CMR tasks, specifically Hit-Rate (HR) and Normalized Discounted Cumulative Gain (nDCG). 

\begin{table*}[t]
    \LARGE 
    \renewcommand{\arraystretch}{1} 
    \centering
    \caption{Performance comparison of different CMR methods}
    \label{tab:mytable}
    \adjustbox{max width=\textwidth}{
    
    \begin{tabular}{cccccccccccccccccccc}
    \toprule
    & & & \multicolumn{2}{c}{\textbf{de}} & \multicolumn{2}{c}{\textbf{jp}} & \multicolumn{2}{c}{\textbf{in}} & \multicolumn{2}{c}{\textbf{fr}} & \multicolumn{2}{c}{\textbf{ca}} & \multicolumn{2}{c}{\textbf{mx}} & \multicolumn{2}{c}{\textbf{uk}} \\
    \cmidrule(lr){4-5} \cmidrule(lr){6-7} \cmidrule(lr){8-9} \cmidrule(lr){10-11} \cmidrule(lr){12-13} \cmidrule(lr){14-15} \cmidrule(lr){16-17}
    & & & nDCG@10 & HR@10 & nDCG@10 & HR@10 & nDCG@10 & HR@10 & nDCG@10 & HR@10 & nDCG@10 & HR@10 & nDCG@10 & HR@10 & nDCG@10 & HR@10 \\
    \midrule
    & GMF & & 0.3123 & 0.4673 & 0.1709 & 0.2936 & 0.4509 & 0.5146 & 0.2813 & 0.4336 & 0.2817 & 0.4308 & 0.5208 & 0.6150 & 0.4506 & 0.5805 \\
    & MLP & & 0.3200 & 0.4829 & 0.1926 & 0.3388 & 0.4539 & 0.5774 & 0.2912 & 0.4683 & 0.3015 & 0.4550 & 0.5362 & 0.6379 & 0.4526 & 0.5813 \\
    & NMF & & 0.3378 & 0.4922 & 0.1938 & 0.3470 & 0.4667 & 0.5356 & 0.2960 &  0.4562 & 0.2867 & 0.4331 & 0.5390 & 0.6230 & 0.4569 & 0.5806 \\
    & MAML & & 0.2808 & 0.4716 & 0.1770 & 0.3555 & 0.4320 & 0.5293 &  0.2785 & 0.4311† &  0.2794 & 0.6764 & 0.5288 & 0.4939† & 0.4296 & 0.6335† \\
    & NMF-FOREC & & 0.2835 & 0.4742 & 0.1758 & 0.3574 & 0.4345 & 0.5447 &  0.2816 & 0.4468 & 0.2772 & 0.4713 &  0.5302 & 0.4673 & 0.4330 & 0.6079 \\
    & MA-GMF & & 0.3140 & 0.4779 & 0.1914 & 0.3388 & 0.4579 & 0.5188 & 0.2777 & 0.4382 & 0.3000 & 0.4465 & 0.5315 & 0.6299 & 0.4491 & 0.5782 \\
    & MA-MLP & & 0.3063 & 0.4736 & 0.1850 & 0.3121 & 0.4268 & 0.5397 & 0.2962 & 0.4753 & 0.3114 & 0.4699 & 0.5294 & 0.6246 & 0.4546 & 0.5912 \\
    & MA-NMF & & 0.3464 & 0.4909 & 0.2083 & 0.3470 & 0.4475 & 0.5564 & 0.3021 & 0.4432 & $\textbf{0.3201}^{*}$ & $\textbf{0.4768}^{*}$ & 0.5478 & 0.6395 & 0.4651 & 0.5856 \\
    & DGRE-GMF & & 0.3148 & 0.4775 & 0.1935 & 0.3326 & 0.4564 & $\textbf{0.5989}^{*}$ & 0.2998 & 0.4700 & 0.3057 & 0.4550 & 0.5379 & 0.6390 & 0.4559 & 0.5808 \\
    & DGRE-MLP & & $\textbf{0.3501}^{*}$ & $\textbf{0.5019}^{*}$ & 0.2216 & $\textbf{0.3532}^{*}$ & 0.4497 & 0.5439 & $\textbf{0.3301}^{*}$ & $\textbf{0.4812}^{*}$ & 0.3125 & 0.4599 & 0.5501 & $\textbf{0.6448}^{*}$ & $\textbf{0.4656}^{*}$ & $\textbf{0.5857}^{*}$ \\
    & DGRE-NMF & & 0.3324 & 0.4711 & $\textbf{0.2294}^{*}$ & 0.3511 & $\textbf{0.4933}^{*}$ & 0.5941 & 0.3214 & 0.4750 & 0.3060 & 0.4558 & $\textbf{0.5509}^{*}$ & 0.6363 & 0.4571 & 0.5692 \\
    \bottomrule
    \end{tabular}
    }
    \vspace{-1em}
    \end{table*}

\subsection{Results and Discussion for Global Experiment}  
Table \ref{tab:mytable} summarizes the outcomes of training a unified recommendation model across all markets. The DGRE model consistently outperforms market-aware (MA) and meta-learning models in most scenarios, with notable strengths in larger markets.  

Against basic models (GMF, MLP, NMF), DGRE achieves superior results in 34 out of 39 cases. DGRE-GMF outperforms GMF in all cases, while DGRE-MLP and DGRE-NMF exceed their counterparts in 11 and 10 cases, respectively. Performance is strongest in larger markets like the UK and Canada, while smaller markets, such as Germany, show mixed results.  

Compared to meta-learning models (MAML, FOREC), DGRE consistently delivers better results across all markets. DGRE-MLP leads in larger markets like Germany, France, and the UK, while DGRE-NMF performs best in Japan, India, and Mexico. In Canada, all DGRE models outperform both MAML and FOREC.  

In comparison with MA models, DGRE achieves better performance in 27 out of 39 scenarios. DGRE-MLP shows strong results across five markets, including the UK, while DGRE consistently excels in Japan. However, MA-NMF performs best in Canada, potentially due to the sensitivity of the parameter $k$ in graph pooling. Overall, DGRE demonstrates robust advantages, particularly in larger and high-performing markets. 
\begin{figure*}[t]
    \centering
    \resizebox{\linewidth}{!}{
        \begin{minipage}[t]{0.55\linewidth}
            \centering
            \includegraphics[width=\linewidth]{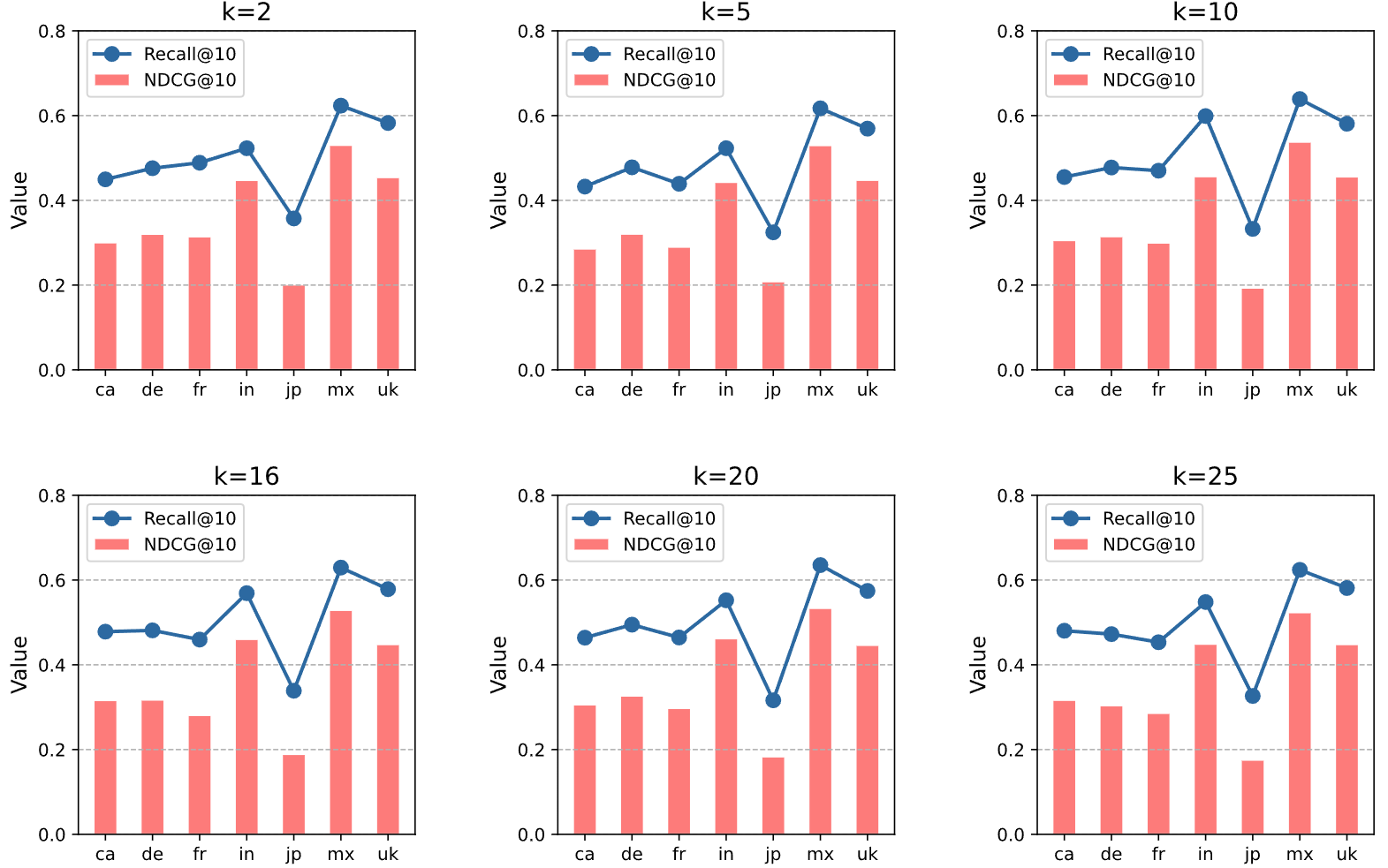}
            \vspace{-2em}
            \caption{Figure for a fixed number of k in different countries.}
            \label{fig:ablation-ncoun1k}
        \end{minipage}
        \hspace{0.02\linewidth}
        \begin{minipage}[t]{0.38\linewidth}
            \centering
            \includegraphics[width=\linewidth]{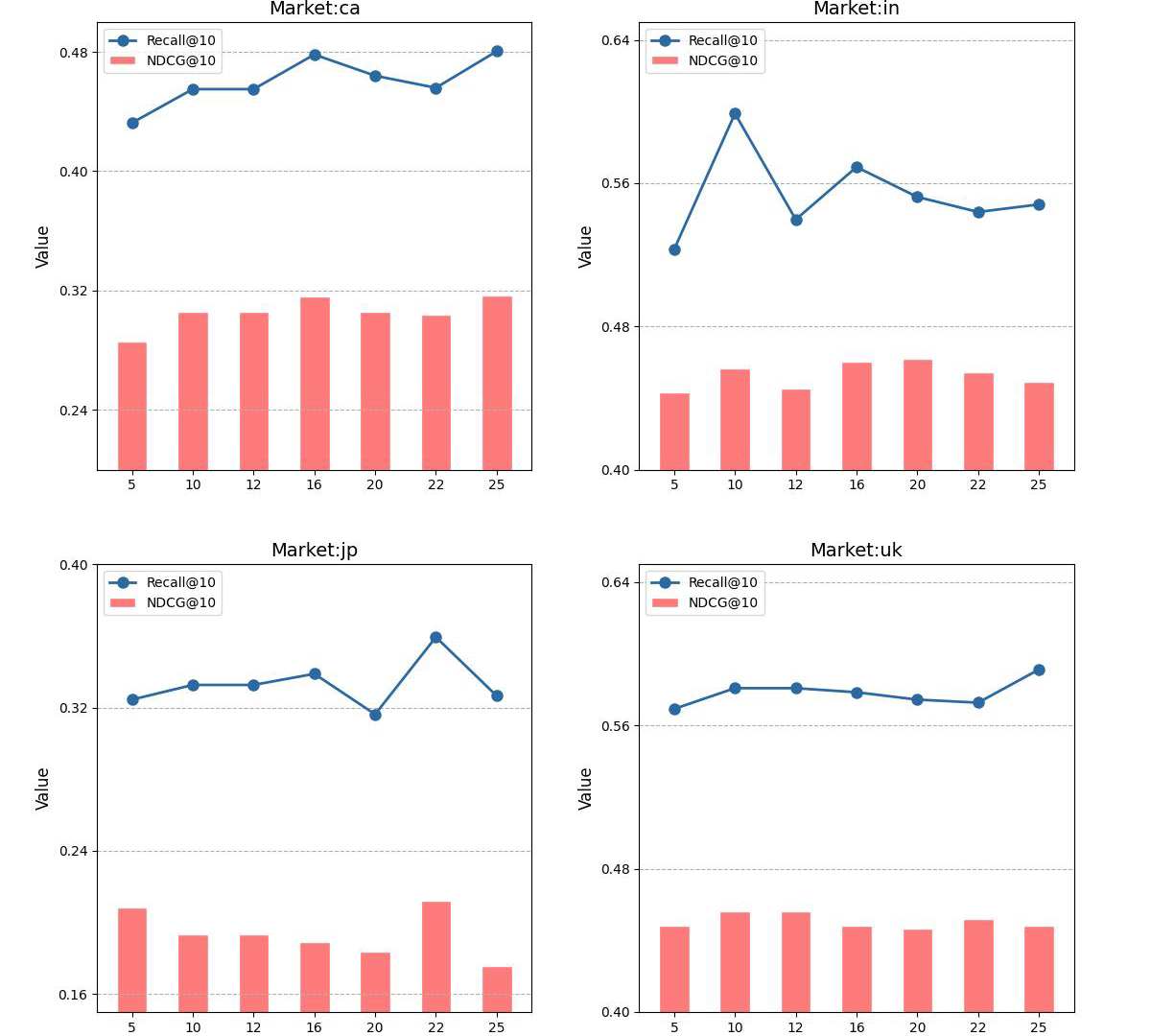}
            \vspace{-2em}
            \caption{Figure for a fixed number of countries in different k).}
            \label{fig:ablation-1counnk}
        \end{minipage}
    }
\vspace{-1.5em}
\end{figure*}

\subsection{Ablation Study on the selection of k influences vector selection}
To evaluate the influence of parameter $k$ on vector selection, we analyzed its effect across various markets and within a fixed market under different $k$ values, as shown in Figures \ref{fig:ablation-ncoun1k} and \ref{fig:ablation-1counnk}. Our results indicate that nDCG@10 and Recall@10 follow consistent trends regardless of the market or $k$ value, with increasing similarity as $k$ converges to a common value.

In individual markets, the optimal $k$ varies. For the 'ca' market, both metrics rise with $k$ and stabilize around $k = 10$. In the 'in' market, the metrics improve steadily, peaking at $k = 10$. The 'jp' market shows stable performance for smaller $k$ values, but variability appears for $k > 10$. The 'uk' market achieves its highest values at $k = 10$ and remains stable beyond this. These results underline the role of $k$ in recommendation performance, offering insights into both market-specific and shared configurations.

\subsection{Ablation Study on the effectiveness of different kinds of embeddings}
To evaluate the impact of market-specific and market-shared prototypes on CMR, we conducted an ablation study with 10 prototype classes by default.
\begin{figure}[htbp]
\centering
    \includegraphics[width=0.4\linewidth]{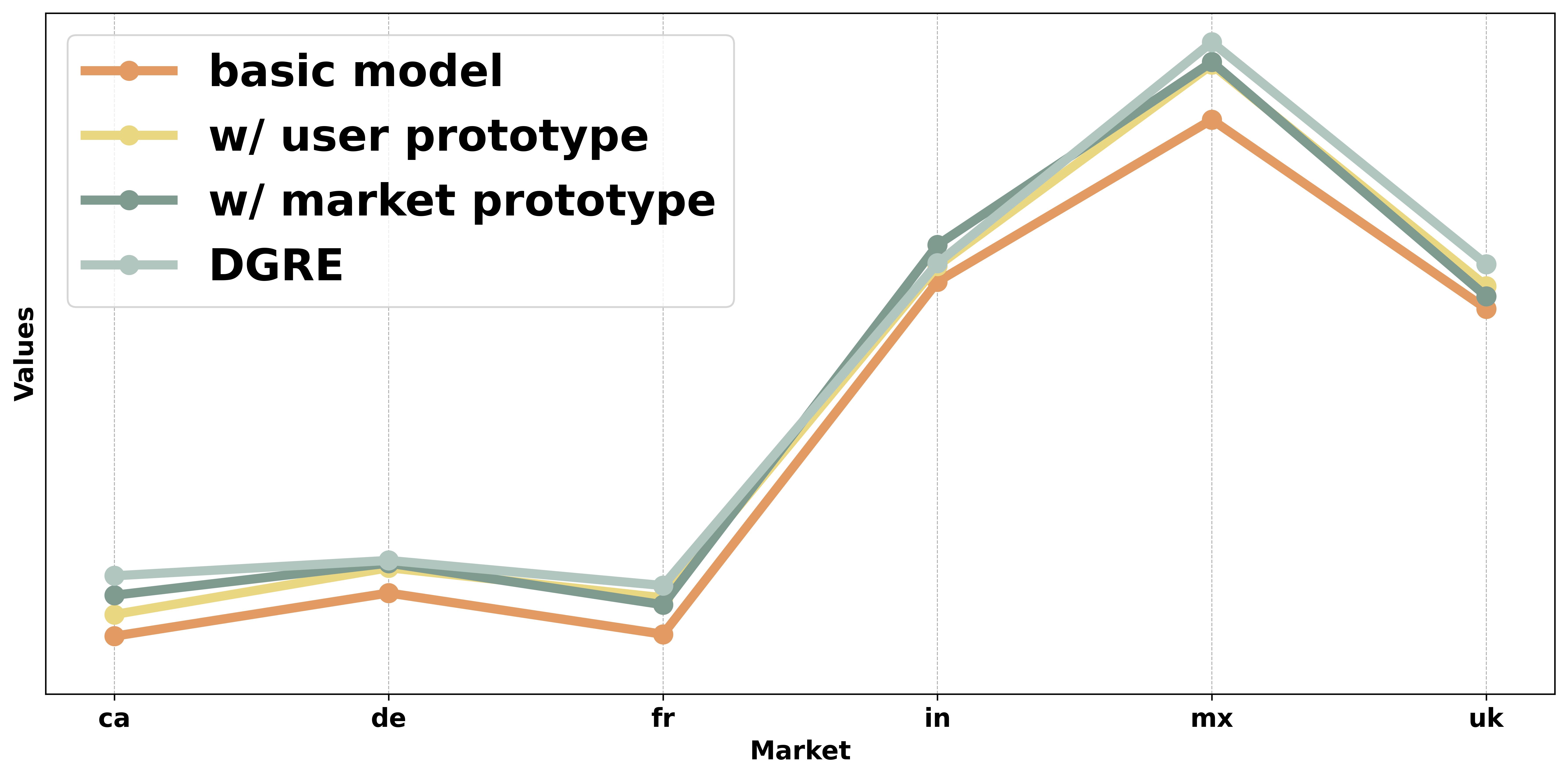} %
    \caption{Figure for different kinds of embeddings}
    \label{fig:ablation-2first}
\end{figure}
As shown in Fig. \ref{fig:ablation-2first}, the "basic model" serves as a reference, providing baseline performance across markets. When market-shared prototypes are introduced, performance remains relatively stable, suggesting their limited effect on CMR. In contrast, when the model operates without prototypes, performance varies across markets. Specifically, the model performs slightly worse in the "de" market but outperforms the basic model in the "fr" market, indicating the benefits of market-specific prototypes.

The DGRE model, which consistently outperforms all others across markets, highlights the substantial improvement achieved by incorporating market-specific prototypes. This study demonstrates that while market-shared prototypes have a modest impact, leveraging market-specific information, as seen in the DGRE model, significantly enhances CMR performance.

\section{Conclusion \& Future Work}
Our research addresses cross-market recommendation (CMR) for promoting multinational goods. Existing methods often ignore shared preferences between markets, limiting generalization. We propose the Dual Prototype Attentive Graph Network (DGRE), which captures both market-specific and shared insights using user and item prototypes. DGRE clusters users to find shared behaviors and aggregates item features for market-specific views. It integrates with existing methods and improves performance on the XMarket dataset. Our work offers a general solution for CMRS, with potential for further improvement through data augmentation and foundation models.
%
%
%
\bibliographystyle{splncs04}
\bibliography{ref}

\end{document}